\documentclass[%
 reprint,
 amsmath,amssymb,
 aps,
]{revtex4-2}

\usepackage{graphicx}
\usepackage{dcolumn}
\usepackage{bm}
\usepackage{xcolor}

\begin{document}




\title{Semiconducting and superconducting properties of 2D hexagonal materials}


\author{Dominik Szcz{\c{e}}{\'s}niak${^1}$}
\email{d.szczesniak@ujd.edu.pl}
\author{Jakub T. Gnyp${^2}$}
\author{Marta Kielak${^1}$}


\affiliation{
${^1}$Institute of Physics, Faculty of Science and Technology, Jan D{\l}ugosz University in Cz{\c{e}}stochowa, 13/15 Armii Krajowej Ave., 42200 Cz{\c{e}}stochowa, Poland\\
${^2}$Condensed Matter Spectroscopy Division, Faculty of Mathematics, Physics and Informatics, University of Gda{\'n}sk, Wita Stwosza 57 Str., 80308 Gda{\'n}sk, Poland
}


\date{\today}


\begin{abstract}

The beginning of high interest in two-dimensional (2D) crystals is marked by the synthesis of graphene, which constitutes exemplary monolayer material. This is due to the multiple extraordinary properties of graphene, particularly in the field of quantum electronic phenomena. However, there are electronic features that are notably missing in this material due to the inherent nature of its charge carriers. Of particular importance is that pristine graphene does not exhibit semiconducting or superconducting properties, preventing related applications. Certain modifications to graphene or even synthesis of sibling materials is needed to arrive with semiconducting and superconducting 2D hexagonal materials. Here, the representative examples of such materials are discussed in detail along with their expected properties. Special attention is given to the unique semiconducting and superconducting phenomena found in these materials {\it e.g.} the non-adiabatic superconductivity, spin- and valley-dependent conductivity or the bulk-like Schottky-type potential barriers. The discussion is supplemented with some pertinent conclusions and perspectives for future work.

\end{abstract}

\maketitle


\section{Introduction}

The discovery of two-dimensional (2D) materials has revolutionized condensed matter physics and materials science, opening new avenues for investigating exotic properties and enabling advanced applications in nanoscale devices \cite{ho2024}. This field gained momentum with the successful isolation of graphene, a single layer of carbon atoms arranged in a hexagonal crystal lattice \cite{novoselov2004electric}, characterized by remarkable electronic features such as ultrahigh carrier mobility, low-energy massless Dirac fermions or the quantum Hall effect \cite{castro2009}.

Despite its remarkable properties, graphene lacks an intrinsic band gap and has a relatively low density of states near Fermi level, limiting its potential for key semiconducting and superconducting applications. These are arguably the two most significant shortcomings of graphene, yet they hold the greatest promise for breakthrough advancements in future 2D electronics \cite{huang2022, uchihashi2017}. To overcome mentioned challenges, modifications to pristine graphene are necessary, often leading to significantly altered or entirely new 2D structures. In this regard, one of the most promising strategies involves introduction of various symmetry-breaking scenarios to tailor electronic properties of 2D materials to specific needs \cite{du2021}.

In this paper, we review recent advancements in the field of 2D hexagonal semiconductors and superconductors, focusing on popular and promising approaches that extend beyond the Dirac approximation. Our analysis emphasizes symmetry-breaking mechanisms associated with the underlying nature of crystal structure and charge carriers. Additionally, we provide general insights that not only lead to meaningful conclusions but also offer perspectives for future research in this field.

\section{Two-dimensional semiconductors}

Bulk semiconductors are the backbone of modern microelectronics and it is expected that their 2D counterparts are set to revolutionize related next-generation applications. The 2D semiconductors are not only poised to miniaturize current microelectronics even further, but also exhibit enhanced gate control and reduced power consumption \cite{duan2024}, high photoconductivity and a strong dielectric screening effect \cite{george2021}, efficient light absorption \cite{nazif2023}, and ion adsorption \cite{ortiz2025}. All highly desirable properties for novel transistors, solar energy converters and other optoelectronic equipment.

Because various modifications of graphene offer either excessively narrow or wide band gaps, other hexagonal monolayers prove to be particularly valid semiconductors \cite{chaves2020}. The most promising materials are 2D transition-metal dichalcogenides (TMDs), which feature mostly direct band gaps of up to approximately $\sim 2$ eV \cite{xie2015}. These semiconducting properties are possible due to the inherently broken inverse symmetry of hexagonal lattice in TMDs and the relatively strong spin-orbit coupling arising from the heavy transition metal atoms \cite{pearce2016}. The latter is also responsible for the two degenerate but in-equivalent local extrema, known as valleys, in either conduction or valence bands, which can be easily electrically manipulated \cite{schaibley2016}, {\it e.g.} with the valley Hall effect. As a result it is possible to encode, process and store information in the degrees of freedom associated with the valleys.

This is to say, the TMDs allows not only for the efficient tunneling currents \cite{szczesniak2016} but also for the charge transport that offers strong valley- and spin-selective coupling with both electric and magnetic fields \cite{szczesniak2020}. Importantly, such transport properties and charge density are heavily dependent on the so-called metal-induced gap states (MIGSs) which mediate both tunneling and charge injection processes \cite{szczesniak2016, szczesniak2020}. In what follows, the valley-spin dependent physics is additionally present in less popular 2D hexagonal semiconductors such as the standalone Janus TMDs \cite{sattar2022} or hetero-structures incorporating wide-gap hexagonal boron nitride ($h$-BN) \cite{zhang2019}.

\begin{figure}[ht!]
\centering
\includegraphics[width=\columnwidth]{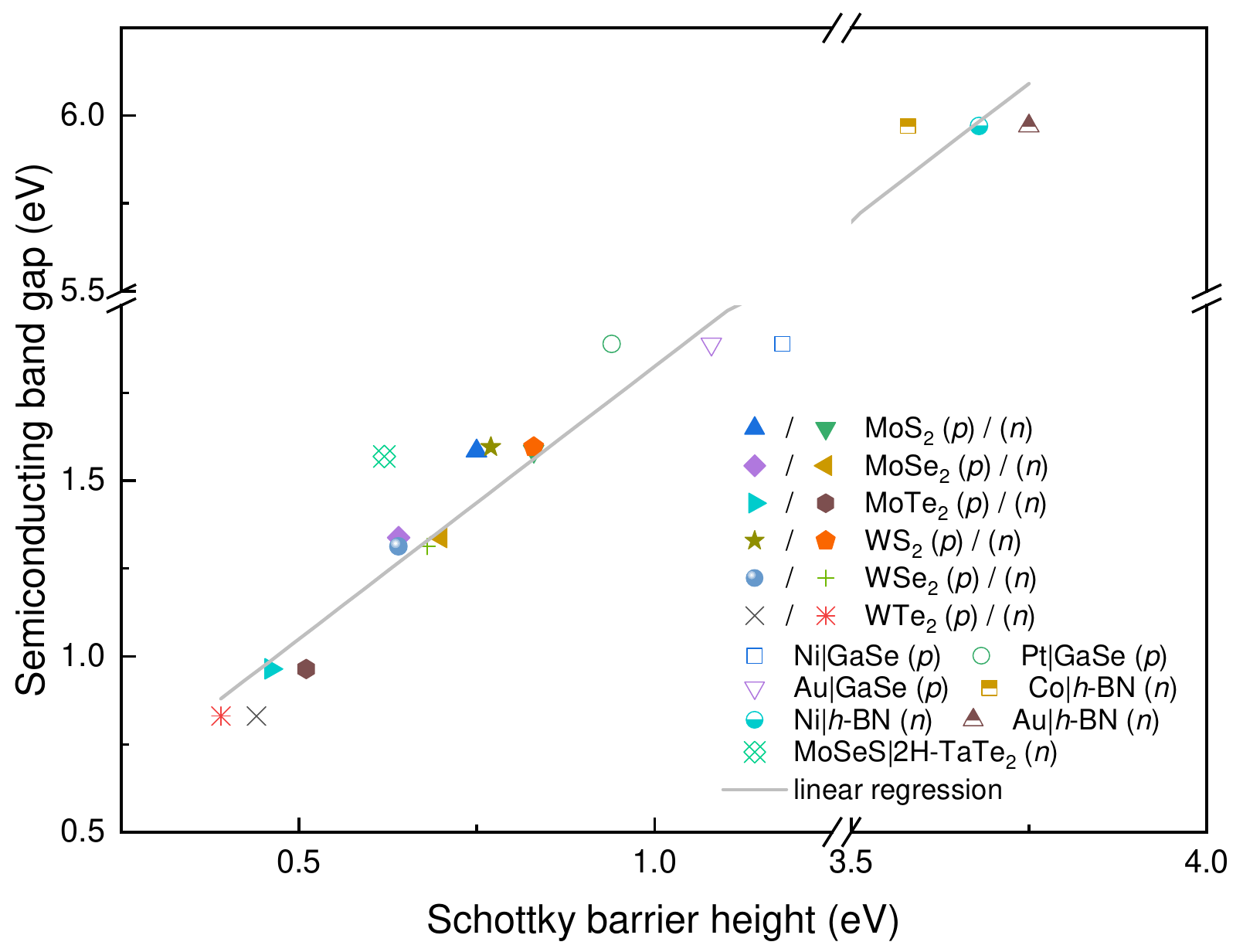}
\caption{The relation between semiconducting band gap and the Schottky barrier height for the selected 2D hexagonal semiconductors. The values are presented for electrons ($n$) and holes ($p$). The data for TMDs, Janus TMDs, GaSe and $h$-BN is adopted from \cite{szczesniak2018}, \cite{zhao2023, picker2025}, \cite{liu2018} and \cite{bokdam2014}, respectively.}
\label{fig01}
\end{figure}

Interestingly enough, the MIGSs appear also to impart locally metallic character of a 2D semiconductors and define charge neutrality level near a metal-semiconductor junction. As such, the MIGSs are argued to contribute to the height of the potential barrier at the metal-semiconductor junction, {\it i.e.} the Schottky barrier height (SBH) \cite{szczesniak2018}. Analysis of this relationship for TMDs yields particularly interesting result. The SBH for the monolayer semiconductors behave virtually the same as for the bulk structures. In details, due to the existence of mentioned MIGSs in the forbidden energy region, the Fermi level of a metal is strongly pinned at the midgap level (see Fig. \ref{fig01}, for more details). Such conclusion is in agreement with not only numerical results \cite{guo2015} but also canonical considerations \cite{szczesniak2018}, explaining little differences in metal-semiconductor electronegativity between cases of different dimensionality.

The above finding can be qualitatively extended on other representative 2D hexagonal semiconductors. In Fig. \ref{fig01}, the SBH of materials sibling to TMDs are depicted in relation to the band gap size. The results suggest that Schottky barriers in these materials may follow the MIGSs model, similarly to the case of TMDs, as the corresponding SBH values fluctuate around the midgap. Note, however, that we plot only those values which present canonical behavior and in the literature there are multiple other instances where SBH deviate from this trend. The reason for that is strong dependence of SBH values on external factors, such as the metal contact type, contact geometries, work function of metal contact, bonding symmetries or crystal field effects \cite{guo2015, szczesniak2018}. Nonetheless, the presented SBH values still reinforce canonical findings, when work function of a metal is relatively high and the junction materials exhibit minimal structural mismatch. This observation is particularly important as it suggests a relatively standardized strategy for engineering Schottky barriers, enabling more efficient injection and transport processes within the discussed class of materials.

\section{Two-dimensional superconductors}

Superconductivity in 2D materials has been extensively studied due to its potential for quantum technologies, low-energy electronics, and novel quantum phases \cite{uchihashi2017}. In this context, of special interest are phonon-mediated superconductors, as they are well-described by the Bardeen–Cooper–Schrieffer (BCS) theory of superconductivity \cite{bardeen1957theory}. Interestingly this conventional picture is not fully preserved since multiple 2D superconductors exhibit comparable phonon and electron energy scales. Consequently, the Migdal's theorem is not satisfied, and additional non-adiabatic effects emerge, significantly influencing the superconducting phase \cite{grimaldi1995nonadiabatic,grimaldi1995nonadiabatic2}.

One of the most important families of conventional 2D superconductors with a hexagonal structure comprises graphene-based systems \cite{uchoa2007superconducting}. Among them, LiC$_6$ stands out as an exemplary material due to its tunable superconducting properties and partial experimental verification \cite{profeta2012phonon, ludbrook2015evidence}. In this material, the superconducting state is induced as a result of chiral symmetry breaking caused by the lithium adatoms that lift Fermi energy near Van Hove singularity \cite{szczesniak2014influence}. Although the superconducting phase exists in bulk LiC$_6$, it is characterized by a vanishingly small $T_C = 0.9$ K. However, transitioning to the monolayer form raises $T_C$ to notable several Kelvins range \cite{profeta2012phonon, szczesniak2014influence}. This is a perfect example of how the properties of materials change when their dimensionality is reduced. Beyond that, this superconducting phase is also intriguing due to the enhanced role of already mentioned non-adiabatic effects \cite{szczesniak2019prdlic}. Notably, the influence of these effects has been found to scale with the strength of superconductivity, suggesting that they play a crucial role in shaping the superconducting properties \cite{szczesniak2023scalability}. Here, the superconducting phase was engineered via strain, a strategy that can be used in other 2D superconductors to improve their properties, {\it e.g.} in silicene \cite{silicene2014}.

Similarly to the above, one can achieve superconductivity in graphene by intercalation with calcium (Ca$_x$C$_6$) \cite{chapman2015superconductivity,ichinokura2016superconducting}. For this structure, the superconducting properties are heavily related to the confinement of the Ca layer and the induced charge carrier concentration \cite{chapman2015superconductivity}. Calculations based on the adiabatic Eliashberg regime show, that Ca layered graphene is a strongly coupled electron-phonon superconductor, where low-energy Ca vibrations are crucial for the pairing mechanism \cite{margine2016electron}. This also suggests the possible impact of non-adiabatic effects, which were revealed in the lithium-intercalated case.

Superconductivity in graphene can also be induced through electron and hole doping \cite{zhou2015high, szczesniak2021non}, where carbon atoms are substituted with foreign dopants that break sublattice symmetry. In both cases, the presence of a shallow conduction band in doped graphene leads to a violation of the Migdal's theorem. In electron-doped graphene, non-adiabatic effects reduce the superconducting transition temperature $T_C$ \cite{szczesniak2021non}, with their influence becoming more pronounced as the depairing Coulomb interaction increases. Similarly, in hole-doped graphene, these effects contribute to the suppression of the superconducting energy gap \cite{Kaczmarek_2023}. However, hole-doped graphene appears more robust against non-adiabatic effects than its electron-doped counterpart, resulting in a relatively smaller reduction in $T_C$ \cite{szczesniak2021non, Kaczmarek_2023}.

Another example of doped graphene is the so-called graphane, a fully hydrogenated hexagonal layer of carbon atoms. This material can reach even higher critical temperature values (up to $90$ K) than previously mentioned materials \cite{savini2010first}. Moreover, this phase highlights the role of quantum confinement in lifting Fermi energy, which is crucial for achieving high $T_C$ superconductivity. Similarly to other graphene-based superconductors, graphane falls into the strong-coupling regime of superconductivity \cite{durajski2015holegrap}. Note, that similar mechanisms could be exploited in other 2D materials, such as hydrogenated silicene or germanene, where doping can enhance electron-phonon coupling, leading to promising superconducting properties \cite{durajski2014estimation}.

Final prominent example is yet another sibling to the LiC$_6$, namely Ca-doped hexagonal boron nitride (Ca-h-BN) \cite{shimada2020theoretical}. In this material, the non-adiabatic effects are once again relatively strong, leading to a strong renormalization of the electron-phonon interaction and reducing the superconducting gap function \cite{drzazga2022breakdown,krok2023critical}. Interestingly, its critical temperature is lower than predicted by conventional Eliashberg theory \cite{drzazga2022breakdown}, suggesting also renormalization of T$_{C}$.

\begin{figure}[ht!]
\centering
\includegraphics[width=\columnwidth]{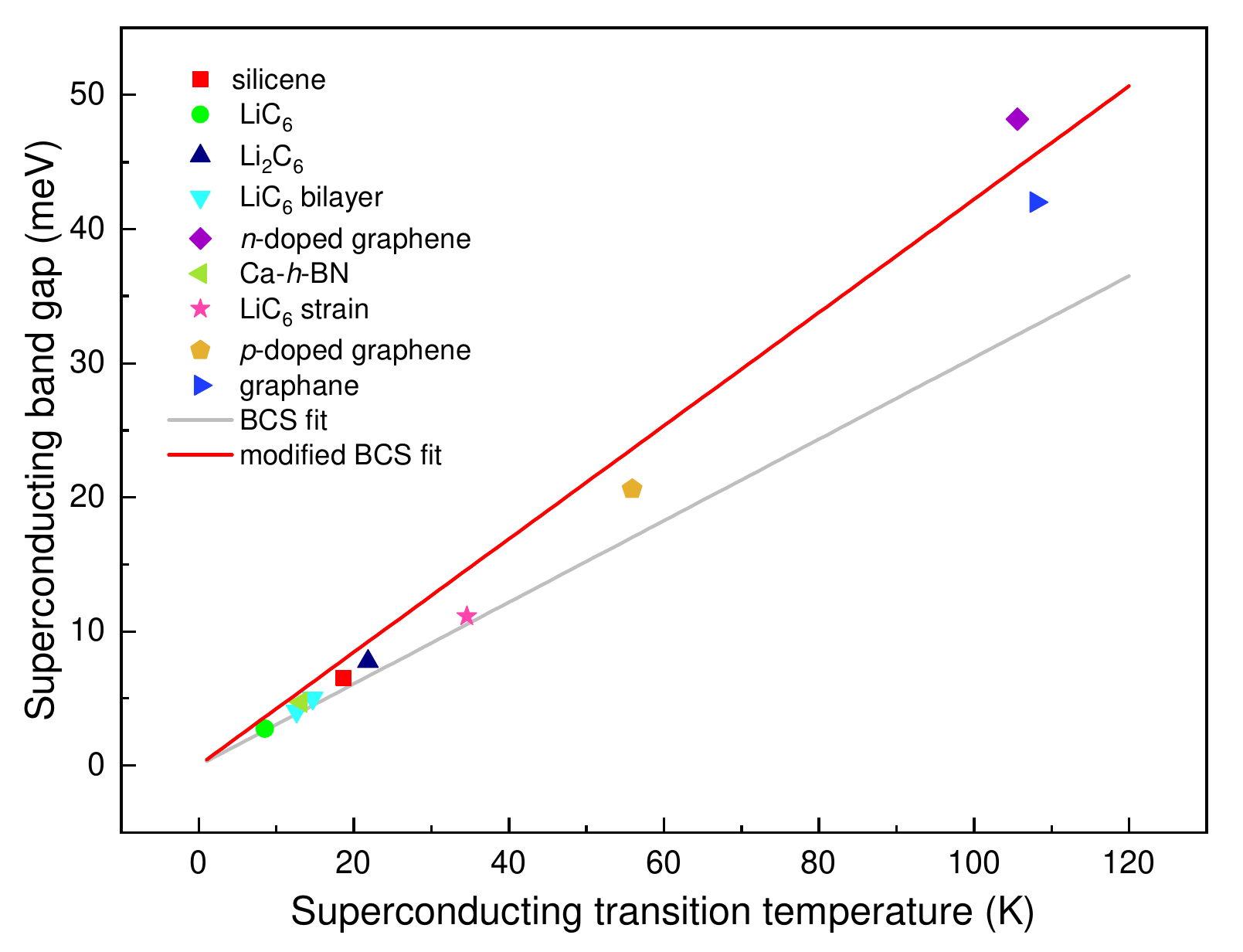}
\caption{The relation between superconducting band gap and superconducting transition temperature in selected 2D hexagonal semiconductors. The data for silicene, LiC$_6$ and Li$_2$C$_6$, LiC$_6$ bilayer, $n$-doped graphene, Ca-$h$-BN, LiC$_6$ strain, $p$-doped graphene and graphane is adopted from \cite{silicene2014}, \cite{szczesniak2014influence}, \cite{szczeniak2015}, \cite{szczesniak2021non}, \cite{drzazga2022breakdown}, \cite{szczesniak2023scalability}, \cite{drzazga2023acta}, and \cite{durajski2015holegrap}, respectively.}
\label{fig02}
\end{figure}

In Fig. \ref{fig02}, the superconducting band gap is given as a function of T$_{C}$, for selected 2D superconductors mentioned in the text. The dependence clearly exhibits a linear increasing trend. However, the data point cannot be fully approximated by the standard BCS fit, $\Delta/k_{B}T_{C}=3.53$, where $\Delta$ is the superconducting band gap and $k_{B}$ denotes Boltzmann constant \cite{bardeen1957theory}. The ratio has to be increased to the value of $\sim 4.9$ to average the data (see modified BCS fit). This suggests increased role of strong-coupling and retardation phenomena. However, as shown by the studies discussed in this paper, more sophisticated methods are required to additionally account for the non-adiabatic effects.

\section{Summary and Conclusions}

In conclusion, the presented overview of 2D hexagonal semiconductors and superconductors highlights their remarkable versatility and technological relevance. The robust electronic properties of semiconducting materials, particularly TMDs, including their gate-tunable valley-spin selectivity and Schottky barrier characteristics, establish them as fundamental building blocks for future nanoelectronics \cite{szczesniak2018,guo2015}. As for the superconducting materials, the discussed findings emphasize the increased role of strong-coupling and non-adiabatic effects, which profoundly influence their properties. Moreover, the discovery of novel superconductors, {\it e.g.} transition-metal monohalides, expands the field into the new directions, paving the way for advancements in high-temperature and strain-engineered superconductivity at low-dimensions \cite{dong2022superconductivity,zhoua2024predicting}. As a summary, continued research in these areas holds great potential for driving significant advancements in quantum technologies, nanoelectronics and energy-efficient devices.

\bibliographystyle{apsrev}
\bibliography{apssamp}

\end{document}